# Zero-temperature d-wave superconducting phase transition

Igor F. Herbut

*Department of Physics, Simon Fraser University, Burnaby, British Columbia,
Canada V5A 1S6*

The Ginzburg-Landau-Wilson theory that describes the disordered metal - d-wave superconductor phase transition at zero temperature is derived at weak coupling. The theory represents an interacting dissipative system of bosonic Cooper pairs in an effective random potential. I show that there exists a wide crossover regime in the theory controlled by a line of Gaussian fixed points, each of which in two dimensions is characterized by a different universal value of the dc critical conductivity. Relation to experiments on overdoped and underdoped cuprates is discussed.

From the point of view of quantum phase transitions [1], [2], [3], the BCS superconducting transition at zero temperature ($T = 0$) represents quite an unusual example. While the thermal superconducting transition in principle falls into the universality class of the XY-model, at $T = 0$ Fermi liquid suffers the BCS instability at an infinitesimal attractive interaction, which completely eliminates pair fluctuations. As long as the non-superconducting state remains a metal, the quantum phase transition into an s-wave superconductor is governed by a Gaussian fixed point, even in the presence of weak non-magnetic impurities and Coulomb interaction [4]. This peculiar situation is caused by the same infrared singularity responsible for the Cooper phenomenon, which here renders the $T = 0$ action a non-local function of the superconducting order parameter (OP), and its Landau expansion ill-behaved. For an s-wave superconductor with weak disorder Anderson theorem guarantees the same singularity still to appear [5], [6] and the nature of the Fermi liquid - superconductor transition essentially remains unchanged by the presence of randomness.

The situation may be radically different if the perturbations are pair-breaking, like in case of a d-wave superconductor with non-magnetic impurities. The effect of disorder on a d-wave superconductor becomes a particularly pressing issue in the context of high-temperature superconductors, where at $T = 0$ the superconducting transition can be tuned by a change of doping or of the level of disorder in the system [7], [8]. Such quantum phase transitions (presumably of superconductor-insulator, or of superconductor-metal type) would correspond to the $T = 0$ end points of the superconducting region in the standard doping-temperature phase diagram of a high-$T_c$ material [9]. In this Letter I obtain the effective action that describes the Fermi liquid - d-wave superconductor $T = 0$ critical point in the limit of weak coupling. Such a theory could be expected to apply to the quantum superconductor-metal transition induced by overdoping a high-$T_c$ material. My starting point is a well-known notion that, when disorder is present, the attractive interaction in the d-wave channel has to be finite to produce pairing even at $T = 0$ [6]. I show that this enables one to derive a *local* $T = 0$ Ginzburg-Landau-Wilson (GLW) action for the fluctuating d-wave OP, which represents a system of interacting dissipative bosons (Cooper pairs) in an *effective* single-particle random potential. The singularities that in the s-wave case yield a non-local $T = 0$ GLW action [4] turn out to be demoted to irrelevant terms (in the renormalization group sense) by the d-wave symmetry of the superconducting state. In a general case without the particle-hole symmetry the pair propagator has both a dissipative and a propagative term linear in frequency. I find that the quantum critical behavior in the GLW theory is governed by a line of Gaussian fixed points (all with same critical exponents $z = 2$ and $\nu = 1/2$), except in an exponentially narrow critical region. In the Gaussian regime the small propagative term in the pair propagator represents a marginal operator, and leads to a *non-universal* singular part of the $T = 0$ dc conductivity at the transition in two dimensions. I conclude by commenting on the experiments on underdoped and overdoped cuprates in light of these results.

Consider the BCS pairing Hamiltonian for spin-1/2 fermions in a standard Gaussian single-particle random potential $V(\vec{r})$, with $\overline{V(\vec{r})V(\vec{r}')} = W\delta(\vec{r} - \vec{r}')$, with the fermions interacting only via weak attractive potential in the d-wave channel (Coulomb repulsion neglected). For simplicity I focus on two dimensions (2D) but neglect the weak-localization effects [10] (see later), so that the normal state is assumed to be a disordered metal. A generalization to dimensions $d > 2$ is then straightforward. Using the standard replica trick [11], the disorder-averaged free energy at $T = 0$ is then $\overline{F} = \lim_{n \to 0}(1 - \overline{Z^n})/n$, with $\overline{Z^n} = \int D[\Psi^\dagger, \Psi] \exp(-S)$, $S = S_0 + S_{int} + S_{dis}$, and:

$$S_0 = \sum_{\vec{q},i,\sigma} \int \frac{d\omega}{2\pi}(i\omega - \xi(\vec{q}))\Psi_{i,\sigma}^\dagger(\vec{q},\omega)\Psi_{i,\sigma}(\vec{q},\omega), \qquad (1)$$

$$S_{int} = -\sum_{\vec{k},\vec{k}',\vec{q},i,\sigma} \int \frac{d\omega d\nu d\nu'}{(2\pi)^3} \frac{V_{\vec{k},\vec{k}'}}{2}\Psi_{i,\sigma}^\dagger(\vec{k}+\vec{q},\nu+\omega)$$

$$\times \Psi_{i,-\sigma}^\dagger(-\vec{k},-\nu)\Psi_{i,-\sigma}(-\vec{k}',-\nu')\Psi_{i,\sigma}(\vec{k}'+\vec{q},\nu'+\omega), \qquad (2)$$

$$S_{dis} = -\frac{W}{2}\sum_{\vec{k}_1...\vec{k}_4,i,j,\sigma,\sigma'} \int \frac{d\omega d\omega'}{(2\pi)^2}\delta(\vec{k}_1+\vec{k}_2-\vec{k}_3-\vec{k}_4)$$

$$\times \Psi_{j,\sigma'}^\dagger(\vec{k}_4,\omega')\Psi_{j,\sigma'}(\vec{k}_2,\omega')\Psi_{i,\sigma}^\dagger(\vec{k}_3,\omega)\Psi_{i,\sigma}(\vec{k}_1,\omega), \qquad (3)$$

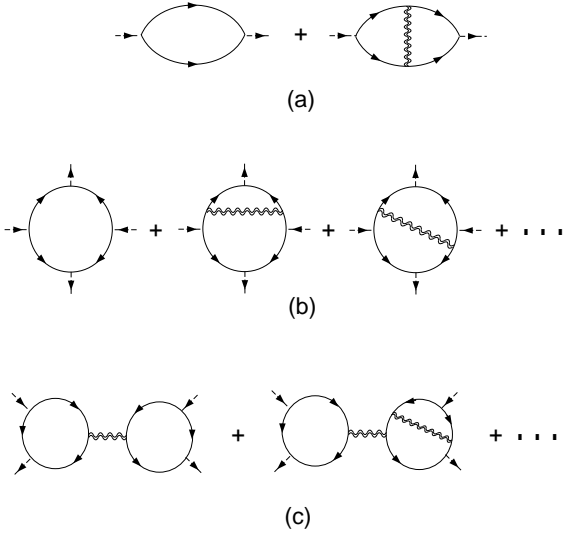

FIG. 1. Quadratic (a), replica-diagonal quartic (b), and replica-off-diagonal quartic (c) kernels in the GLW theory.

with $\Psi$s being Grassman variables. Index $i = 1, ...n$ numerates replicas, $\sigma$ spin, and integrations over the continuous Matsubara frequencies are over the whole real axis. I work in units in which $\hbar = k_B = 1$, and assume $V_{\vec{k},\vec{k}'} = V\cos 2\theta_{\vec{k},\vec{k}'}$, with $V > 0$ (attraction), and $\theta_{\vec{k},\vec{k}'}$ the angle between the two momenta. $\xi(\vec{q}) = (q^2/2m) - \epsilon_f$, where $\epsilon_f$ is the Fermi energy, and I neglect the dependence of the pairing interaction on amplitudes of the momenta. The action $S$ represents a minimal model that contains both the d-wave superconducting instability and the random potential.

The standard procedure [2] is to decouple $S_{int}$ in the Cooper channel with a complex Hubbard-Stratonovich field

$$\Delta(\vec{k}, \vec{q}, \omega) = V^{1/2}\Delta(\vec{q}, \omega)\cos 2\theta_{\vec{k}}, \qquad (4)$$

and then integrate out fermions, to be left with the functional integral over $\Delta(\vec{q},\omega)$ only. If performed exactly, this would amount to exact rewriting of the partition function. Since this is practically impossible, it is useful to expand the resulting action in powers of the OP field

$$S_{GLW}[\Delta] = \sum_{1,i} K_2(1)\Delta_i^*(1)\Delta_i(1) + \sum_{1,2,3,i,j} K_{4,ij}(1,2,3)$$
$$\times \Delta_i^*(1)\Delta_i(2)\Delta_j^*(3)\Delta_j(1+3-2) + O(\Delta^6), \qquad (5)$$

where I shortened the notation $\{\vec{q}_1, \omega_1\} \to 1$, and dropped the regular $\Delta = 0$ part of the action. The quartic kernel in principle has replica-diagonal and off-diagonal parts: $K_{4,ij} = K_4^{diag}\delta_{ij} + K_4^{off}$. The quantum-mechanical average of the fluctuating field $\Delta$ is proportional to the d-wave superconducting OP. With disorder the integration over fermions can not be performed exactly even for the $\omega = q = 0$ mode. The kernels $K_2$, $K_4$, ... in the GLW action $S_{GLW}[\Delta]$, however, are determined exclusively by the connected correlation functions of the disordered metal, and may be represented diagrammatically as in Fig. 1. Full lines stand for the fermion propagator dressed by the random potential, and wavy lines for the fully renormalized vertex $W_{ren}$. I assume that the fermion Green's function is diagonal in replica indices and given by the standard expression for a disordered metal

$$G^{-1}(\vec{q}, \omega) = i\omega - \xi(\vec{q}) + (i/2\tau)\mathrm{sign}(\omega). \qquad (6)$$

At a weak attraction, the phase transition will occur at weak disorder, so I take the scattering time in the self-consistent Born approximation, $\tau = (2\pi W\mathcal{N})^{-1}$, with $\mathcal{N} = m/2\pi$ being the constant density of states per spin in $d = 2$. Accordingly, the vertex is approximated by the sum of particle-particle ladders (Fig. 2), which at small frequencies and momenta yields:

$$W_{ren}(Q, \Omega, \nu) = \frac{W}{|2\nu + \Omega|\tau + (l^2Q^2/2)}. \qquad (7)$$

The choice of ladder diagrams for the vertex ensures that for an s-wave superconductor one reproduces a divergent $K_2 \propto -V\mathcal{N}\ln(lq/2)$ when $q \to 0$ (at $\omega = 0$) [4]. This guarantees that in the s-wave case the system at $T = 0$ becomes unstable towards pair formation at an infinitesimal attraction, in accord with the Anderson theorem [5]. Within the ladder approximation, for the uniform OP mode the present calculation of $K_2$ becomes equivalent to the standard mean-field approach [6]. To the lowest order in $q$ and $\omega$, the quadratic (Fig. 1a), and the replica-diagonal part of the quartic kernel (at zero external momenta and frequencies) (Fig. 1b) in the $T = 0$ action for the OP field are then

$$K_2(\vec{q}, \omega) = 1 - g\ln(2\epsilon_f\tau) + g(\frac{ql}{2})^2 +$$
$$g\tau(|\omega| + i\frac{\delta}{4\epsilon_f\tau}\omega) + O(\omega^2, q^4\ln(ql)), \qquad (8)$$

$$K_4^{diag}(0) = \frac{5g^2\tau^2}{12\mathcal{N}}, \qquad (9)$$

where $g = V\mathcal{N}/2$, and to regularize the standard ultraviolet divergence in $K_2(0)$ I employed a high-energy cutoff $\Omega \approx \epsilon_f \gg 1/\tau$. $l = v_f\tau$ is the elastic mean-free path. The parameter $\delta = (\epsilon_f - \Omega)/\epsilon_f$ measures the particle-hole asymmetry in the model.

Few remarks on the derivation of the above results are in order. A great simplification that arises due to the d-wave symmetry of the OP is that in ladder approximation for the vertex the contributions of second diagrams in Figs. 1a and 1b at zero external momenta vanish due to integrations over the angles of momenta in the loops. The leading frequency dependence of $K_2$ in this approximation is thus entirely determined by the particle-particle bubble at Fig. 1a. At $q \neq 0$, however, the insertion of the particle-particle ladder into the second diagram at Fig. 1a gives a non-analytic momentum dependence $\propto q^4 \ln(ql)$ (at $\omega = 0$). This is, however, irrelevant at long-distances compared to the already present and regular $q^2$ term. This again is a consequence of the assumed d-wave symmetry of the OP. The reader should recall that in case of an s-wave superconductor exactly the same ladder approximation for the vertex leads to a diverging $K_2$ when $q \to 0$, as noted earlier. Finally, in the calculation of $K_4(0)$ the third diagram in Fig. 1b is finite and actually of the same order in $W$ as the first, and is retained in the Eq. 9. Within the ladder approximation all the higher-order diagrams for $K_4^{diag}(0)$ vanish exactly.

An important new element brought by disorder is the replica-off-diagonal contribution to the quartic kernel given by the sum of goggles-like diagrams at Fig. 1c. Besides mixing replicas, it conserves frequency for each replica separately, similarly as in $S_{dis}$ in the fermionic action. In our ladder approximation for $W_{ren}$ the second and all higher-order diagrams in Fig. 1c again vanish for zero external momenta, and one obtains:

$$K_4^{off-diag}(0) = -2g^2 W \tau^2 F(k_f l)(1 + O(\delta/\epsilon_f \tau)), \quad (10)$$

where $F(x) \approx (3\pi/8) \ln(x)/x$, for $x \gg 1$ (weak-disorder limit). Rescaling the OP field to bring the coefficient of the $|\omega|$ term in Eq. (8) to unity as $\Phi = (g\tau)^{1/2}\Delta$, and retaining only the leading terms in the gradient expansion (Eqs. (8), (9) and (10)), the GLW action for the fluctuating d-wave OP at $T = 0$ in real space takes a suggestive form:

$$S[\Phi] = \sum_i \int d^2\vec{r} d\tau [\Phi_i^*(\vec{r}, \tau)(|\partial_\tau| + (\delta/(4\epsilon_f \tau))\partial_\tau)\Phi_i(\vec{r}, \tau)$$
$$+ \frac{1}{2m_b}|\nabla \Phi_i(\vec{r}, \tau)|^2 - \mu_b|\Phi_i(\vec{r}, \tau)|^2 + \lambda|\Phi_i(\vec{r}, \tau)|^4]$$
$$- \frac{W_b}{2}\sum_{i,j}\int d^2\vec{r} d\tau d\tau' |\Phi_i(\vec{r}, \tau)|^2 |\Phi_j(\vec{r}, \tau')|^2, \quad (11)$$

where by the operator $|\partial_\tau|$ one means $|\omega|$ in Matsubara space. This GLW action describes a system of bosons, with a mass-like parameter $m_b = 2\tau/l^2$, a "chemical potential" $\mu_b = (\ln(2\epsilon_f \tau) - g^{-1})/\tau$, and the delta-function repulsion of strength $\lambda = 5/(12\mathcal{N})$, coupled to a bath of gapless fermions. The last, replica-coupling, term in the Eq. (11) is precisely as if it derived from averaging of the boson free-energy with a single-particle random potential term $\int V_b(\vec{r})|\Phi(\vec{r}, \tau)|^2$ in the action, with $\overline{V_b(\vec{r})V_b(\vec{r'})} = W_b \delta(\vec{r} - \vec{r'})$, where $W_b = 4W F(k_f l)$.

The obtained form of the $T = 0$ GLW action appears natural and likely to be independent of the details of the underlying electronic Hamiltonian. Finite scattering rate in the fermion Green's function (6) cuts off the infrared singularities in case of d-wave and regularizes the theory very much like the finite temperature would do [12]. Thus a local GLW theory. Besides the standard frequency and momentum dependent terms, the pair propagator contains also a dissipative term of the Caldeira-Leggett form, signifying that the pairs can decay into gapless fermions. Notice that the ratio of the coefficients of the dissipative ($|\omega|$) and the propagative ($i\omega$) terms is proportional to dimensionless conductivity of the normal phase, so that for a good metal $\epsilon_f \tau \gg 1$ and the propagative term is negligible, similarly to the situation at finite-$T$ at weak coupling [12]. The repulsion between the pairs, $\lambda$, is proportional to the average spacing between the electron's energy levels near $\epsilon_f$, since its origin is in the Pauli principle. Finally, the bosonic pairs themselves see an effective random potential, which derives from the underlying disorder for electrons.

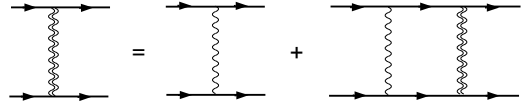

FIG. 2. Particle-particle ladder for the vertex $W_{ren}$. The sum of incoming momenta is $\vec{Q}$, frequency of the upper lines is $\nu + \Omega$, and of the lower lines $-\nu$.

The critical behavior of the interacting theory (11) with both dissipation and disorder is not known, and poses a rather challenging problem. In the weak-coupling (or weak-disorder) limit, however, the effective disorder $W_b$ seen by the bosons matters only in a narrow critical region, and can be neglected in the first approximation. To see this, let me first drop the propagative term in the action when $\epsilon_f \tau \gg 1$. The singular part of the disorder-averaged free-energy of the bosonic system described by the action (11) depends then on two dimensionless couplings: $\hat{\lambda} = \lambda m_b$, and $\hat{W}_b = W_b m_b/|\tilde{\mu}_b|$, where $\tilde{\mu}_b$ is the renormalized chemical potential for the pairs. Using the above results for the bosonic parameters, it follows that $\hat{W}_b$ becomes comparable to $\hat{\lambda}$ only when $|g - g_c|/g_c \approx \exp(-1/g_c)$, where $g_c \ll 1$ is the critical coupling. At weak enough coupling one is thus provided with a wide crossover region where the last term in GLW (11) can be safely neglected.

The system described by the GLW theory (11) with $W_b = 0$ in general has two phases at $T = 0$: the condensed phase, i. e. a d-wave superconductor, at $\tilde{\mu}_b > 0$, and the normal phase at $\tilde{\mu}_b < 0$, with the quantum transition from one phase into another corresponding to pair-breaking. I assumed the normal phase to be a disordered metal. Right in 2D the weak-localization corrections will turn the normal state into an Anderson insulator at distances of the order of localization length $L_{loc} \propto l \exp(k_f l)$ [10]. In the replica formalism this effect can be included by coupling the OP to massless "diffusons" [11], i. e. rotations of the elec-tron self-energy in replica space which leave the action at $\Delta = 0$ invariant. At weak coupling, however, these effects can become significant only in an even narrower region $|g - g_c|/g_c \propto g_c \exp(-\exp(1/g_c))$, i. e. when the superconducting correlation length becomes of order of $L_{loc}$. In the Gaussian crossover regime under consideration the weak-localization effects are thus completely negligible even in 2D. The only remaining dimensionless coupling $\hat{\lambda} = 5\pi/(6\epsilon_f \tau) \ll 1$ is irrelevant in all $d \geq 2$, since power counting yields the dynamical critical exponent $z = 2$, and the system is above (or in 2D right at) its upper critical dimension. The correlation length critical exponents in the crossover region is therefore a Gaussian $\nu = 1/2$, and the usual logarithmic corrections to scaling should be expected in 2D. Assuming that the $T = 0$ superconductor-metal transition in overdoped cuprates is described by the GLW theory (11), this, for example, implies that $T_c \propto (x_c - x)^{z\nu} \propto (x_c - x)$ for $x \approx x_c$, where $x_c$ is the critical value of doping. Similarly, the superfluid density should be $\rho_s \propto (x_c - x)^{\nu(d+z-2)} \propto (x_c - x)^{3/2}$, in $d = 3$. This gives $\rho_s \propto T_c^{1+(1/z)}$, in $d = 3$, which in principle is testable experimentally.

Since the normal state here is assumed to be a metal, at $T = 0$ the dc conductivity consists of a finite regular term and a singular contribution: $\sigma = \sigma_r + \sigma_s$, where at weak disorder $\sigma_r = (2\epsilon_f \tau)e^2/h$. Since the critical point at $W_b = 0$ is Gaussian, one can explicitly compute the singular part of the $T = 0$ dc conductivity using the GLW action (11) [13]. Right at the transition in 2D I find:

$$\sigma_s^* = f(\frac{\delta}{4\epsilon_f \tau})\frac{(2e)^2}{h}, \qquad (12)$$

where $f(0) = (\pi^2 - 4)/(16\pi)$, or $0.117(2e)^2/h$ [14]. For $x << 1$, $f(x) = f(0)(1 - ((12-\pi^2)/(\pi^2-4))x^2 + O(x^4))$, and for $x >> 1$, $f(x) = 1/(4x) + O(1/x^2)$. Remarkably, if the system is not particle-hole symmetric, the universality of the singular part of the conductivity [13] at the transition in 2D is lost because the propagative term represents a marginal operator at the Gaussian fixed point. The function $f(x)$ interpolates between the universal values for the particle-hole symmetric theory ($\delta = 0$) and for the theory without dissipation ($\epsilon_f \tau \to 0$), when $\sigma_s^* = 0$. Although (12) is not the directly experimentally relevant conductivity, where typically one assumes zero-frequency and low temperature [15], [16], it is the simplest one to calculate and may be indicative of the general situation. One could expect a non-universal $\sigma_s^*$ whenever there are both linear frequency terms present in the OP action. If such a theory happens to be describing a superconductor-insulator transition, the total $\sigma = \sigma_s$ could still remain non-universal at the transition, much like what happens in s-wave superconductors in magnetic field [19].

The effective disorder $W_b$ in (11) is a relevant perturbation at the Gaussian fixed point for all $d < 4$. The critical region dominated by $W_b$ also grows with increase of disorder. At a larger disorder, if the trend found here persists, the relative importance of the propagative and the dissipative terms in the OP action could become reversed. For a bad metal, if the coefficient of the dissipative term becomes small compared to the propagative term the GLW theory (11) would approach the action for dirty-bosons [17], [18]. It is not inconceivable that at strong disorder the underdoped cuprates may be describable by the GLW theory in the form of (11), except with a small dissipative term, which would make the phase transition closer to the superfluid-Bose glass universality class. This would account naturally for the universal critical conductivity observed at the superconductor-insulator transition in $Zn$-substituted underdoped cuprates [7].

I thank to Professors S. Sachdev, S. Sondhi and Z. Tesanovic for useful discussions and Dr. M. Amin for a collaboration in an early stage of this project. This work has been supported by NSERC of Canada. I am also grateful to the Aspen Center for Physics where a part of this work was carried out.